\documentclass[amsmath,amssymb,showpacs,twocolumn]{revtex4}
\usepackage{graphicx}
\usepackage{dcolumn}
\usepackage{bm}
\makeatletter \@addtoreset{equation}{section} \makeatother

\def\x{\mathbf x}

\begin{document}
\title{No Stability Switching at Saddle-Node Bifurcations of Solitary Waves in
Generalized Nonlinear Schr\"odinger Equations}
\author{Jianke Yang }
\affiliation{%
Department of Mathematics and Statistics, University of Vermont,
Burlington, VT 05401
}%

\begin{abstract}
Saddle-node bifurcations arise frequently in solitary waves of
diverse physical systems. Previously it was believed that solitary
waves always undergo stability switching at saddle-node
bifurcations, just as in finite-dimensional dynamical systems. Here
we show that this is not true. For a large class of generalized
nonlinear Schr\"odinger equations with real or complex potentials,
we prove that stability of solitary waves does not switch at
saddle-node bifurcations. This analytical result is confirmed by
numerical examples where both soliton branches are stable at
saddle-node bifurcations.
\end{abstract}


\maketitle

Saddle-node bifurcation is a well known phenomenon in
finite-dimensional dynamical systems \cite{Guck_Holm}. In this
bifurcation, there are two fixed-point branches on one side of the
bifurcation point and no fixed points on the other side, and the
stability of these two fixed-point branches switches at the
bifurcation point (one branch stable and the other branch unstable).
In nonlinear partial differential equations (which can be viewed as
infinite-dimensional dynamical systems), this bifurcation exists as
well (it is also called fold bifurcation in the literature). For
instance, solitary waves in nonlinear physical systems often exhibit
this type of bifurcation. Examples include the Boussinesq equations
and the fifth-order Korteweg–-de Vries equation in water waves
\cite{Champneys_1996,Akylas_1997,Chen_2000}, the Swift-Hohenberg
equation in pattern formation \cite{Burke_2007}, the nonlinear
Schr\"odinger (NLS) equations with localized or periodic potentials
in nonlinear optics and Bose-Einstein condensates
\cite{Panos_2005,Kapitula_2006,Sacchetti_2009,Akylas_2011}, and many
others. Motivated by stability switching of saddle-node bifurcations
in finite-dimensional dynamical systems, it is widely believed that
in nonlinear partial differential equations, stability of solitary
waves also always switches at saddle-node bifurcations (see
\cite{Panos_2005,Burke_2007,Kapitula_2006,Sacchetti_2009} for
examples). This belief is very pervasive since no counterexample has
been reported yet.

In this paper, we show that this belief of universal stability
switching at saddle-node bifurcations in nonlinear partial
differential equations is incorrect. Specifically, we show that in a
wide class of generalized NLS equations with real or complex
potentials, stability of solitary waves actually does not switch at
saddle-node bifurcations. This fact is proved analytically by using
general conditions of saddle-node bifurcations and
eigenvalue-bifurcation analysis. It is also verified numerically by
several examples, where both branches of solitary waves are stable
at saddle-node bifurcations.

We consider general nonlinear Schr\"odinger-type equations with
arbitrary forms of nonlinearity and external potentials in
multidimensions,
\begin{equation}  \label{e:U}
iU_t+\nabla^2 U+F(|U|^2, \x)U=0,
\end{equation}
where $\nabla^2$ is the Laplacian in the $N$-dimensional space
$\textbf{x}=(x_1, x_2, \cdots, x_N)$, and $F(\cdot, \cdot)$ is a
general function which contains nonlinearity as well as external
potentials. These equations include the Gross-Pitaevskii equation in
Bose-Einstein condensates and nonlinear light-transmission equations
in localized or periodic potentials as special cases
\cite{BEC,Kivshar_book,Yang_SIAM}. Below, we will first focus on the
case where the function $F$ is real-valued, which applies when the
system (\ref{e:U}) is conservative. Extension to the
non-conservative case of complex functions of $F$ will be considered
afterwards.

When the function $F$ is real, Eq. (\ref{e:U}) admits stationary
solitary waves of the form
\begin{equation}  \label{e:Usoliton}
U(\x,t)=e^{i\mu t}u(\x),
\end{equation}
where $u(\x)$ is a localized real function satisfying
\begin{equation}  \label{e:u}
\nabla^2u-\mu  \hspace{0.01cm} u+F(u^2, \x)u=0,
\end{equation}
and $\mu$ is a real propagation constant which is a free parameter.
Under certain conditions, these solitary waves undergo saddle-node
bifurcations at special values of $\mu$
\cite{Panos_2005,Kapitula_2006,Sacchetti_2009,Akylas_2011}. A
signature of these bifurcations is that on one side of the
bifurcation point $\mu_0$, there are no solitary wave solutions; but
on the other side of $\mu_0$, there are two distinct solitary-wave
branches which merge with each other at $\mu=\mu_0$. To derive
conditions for these bifurcations, we introduce the linearization
operator of Eq. (\ref{e:u}),
\begin{equation}  \label{e:L1}
L_1=\nabla^2-\mu+\partial_u[F(u^2, \x)u].
\end{equation}
We also introduce the standard inner product of functions $\langle
f, g\rangle  = \int_{-\infty}^\infty f^*(\x) \hspace{0.05cm} g(\x)
\hspace{0.03cm} d \x$, where the superscript `*' represents complex
conjugation. Our analysis starts with the basic observation that, if
a bifurcation occurs at $\mu=\mu_0$, by denoting the corresponding
solitary wave and the linearization operator as
\begin{equation}
u_0(\x)=u(\x; \mu_0), \quad L_{10}=L_1|_{\mu=\mu_0,\ u=u_0},
\end{equation}
then the linear operator $L_{10}$ should have a discrete zero
eigenvalue. This is a necessary condition for all types of
bifurcations. To derive sufficient conditions for saddle-node
bifurcations, let us assume that this zero eigenvalue of $L_{10}$ is
simple, which is the case for generic bifurcations in one spatial
dimension as well as for many bifurcations in higher spatial
dimensions. Under this assumption, we denote the unique discrete
(localized) eigenfunction of $L_{10}$ at the zero eigenvalue as
$\psi(\x)$, i.e.,
\begin{equation} \label{e:L10psi}
L_{10}\psi=0.
\end{equation}
Since $L_{10}$ is a real operator, we can normalize the
eigenfunction $\psi$ to be a real function. We also denote
\begin{equation} \label{n:G}
G(u;\x)=F(u^2;\x)u, \quad G_k(\x)=\partial_u^k G|_{u=u_0},
\end{equation}
where $k=1, 2, 3, \cdots$. Then the sufficient condition for
saddle-node bifurcations of solitary waves is given by the following
theorem.

\vspace{0.3cm} \textbf{Theorem 1} \ Under the above assumption and
notations, if $\langle u_0, \psi\rangle \ne 0$ and $\langle G_2,
\psi^3\rangle \ne 0$, then a saddle-node bifurcation of solitary
waves occurs at $\mu=\mu_0$ in Eq. (\ref{e:u}).

\vspace{0.2cm} \textbf{Proof. } Solitary waves which exist near
$\mu=\mu_0$ admit the following perturbation series expansions
\begin{eqnarray}
& u(\x; \mu) = \sum_{k=0}^\infty (\mu-\mu_0)^{k/2}u_k(\x).
\label{e:uexpand1}
\end{eqnarray}
Inserting this expansion into Eq. (\ref{e:u}), we get the following
equations for $u_k$ at order $(\mu-\mu_0)^{k/2}$, $k=0, 1, 2,
\dots$:
\begin{eqnarray}
&&\nabla^2u_0-\mu_0 u_0+F(u_0^2, \x)u_0=0,    \label{e:u0} \\
&& L_{10}u_1=0,    \label{e:u1}  \\
&& L_{10}u_2=u_0-G_2u_1^2/2! \hspace{0.05cm},   \label{e:u2}
\end{eqnarray}
and so on. Eq. (\ref{e:u0}) for $u_0$ is satisfied automatically
since $u_0$ is a solitary wave at $\mu=\mu_0$. The $u_1$ solution to
Eq. (\ref{e:u1}) is found from (\ref{e:L10psi}) as
\begin{equation}  \label{s:u1}
u_1=b_1\psi,
\end{equation}
where $b_1$ is a constant. The $u_2$ function satisfies the linear
inhomogeneous equation (\ref{e:u2}). Due to the Fredholm Alternative
Theorem and the fact that $L_{10}$ is self-adjoint, Eq. (\ref{e:u2})
admits a localized solution for $u_2$ if and only if the homogeneous
solution $\psi$ is orthogonal to the inhomogeneous term, i.e.,
\begin{equation}  \label{e:orth1}
\langle \psi, u_0-G_2u_1^2/2 \rangle=0.
\end{equation}
Inserting the solution (\ref{s:u1}) into this orthogonality
condition and recalling the conditions in Theorem 1, we find that
\begin{equation}  \label{e:c12}
b_1=b_1^\pm \equiv \pm \sqrt{\frac{2\langle u_0,
\psi\rangle}{\langle G_2, \psi^3\rangle}}.
\end{equation}
Thus, we get two $b_1$ values $b_1^\pm$ which are opposite of each
other. Inserting the corresponding $u_1$ solutions (\ref{s:u1}) into
(\ref{e:uexpand1}), we then get two perturbation-series solutions of
solitary waves $u(\x; \mu)$ as
\begin{equation}  \label{s:upm}
u^\pm (\x; \mu)=u_0(\x)+b_1^\pm (\mu-\mu_0)^{1/2}\
\psi(\x)+O(\mu-\mu_0).
\end{equation}
If $\langle u_0, \psi\rangle$ and $\langle G_2, \psi^3\rangle$ have
the same sign, then $b_1^\pm$ are real. Recalling that $u_0(\x)$ and
$\psi(\x)$ are both real as well, we see that these
perturbation-series solutions (\ref{s:upm}) give two real-valued
(legitimate) solitary waves when $\mu>\mu_0$, but these solitary
waves do not exist when $\mu<\mu_0$. On the other hand, if $\langle
u_0, \psi\rangle$ and $\langle G_2, \psi^3\rangle$ have the opposite
sign, $b_1^\pm$ are purely imaginary. In this case, the perturbation
series (\ref{s:upm}) give two real-valued solitary waves when
$\mu<\mu_0$ but not when $\mu>\mu_0$.

The above perturbation calculations can be continued to higher
orders. We can show that the two real solitary-wave solutions
(\ref{s:upm}), which exist on only one side of $\mu=\mu_0$, can be
constructed to all orders of $(\mu-\mu_0)^{1/2}$. In addition, these
two solitary waves $u^\pm (\x; \mu)$ merge with each other when
$\mu\to \mu_0$. We can also show that except these two solitary-wave
branches, there are no other solitary-wave solutions near the
bifurcation point. Thus a saddle-node bifurcation occurs at
$\mu=\mu_0$. This completes the proof of Theorem 1.

Stability properties of solitary waves near saddle-node bifurcations
is an important issue. In finite-dimensional dynamical systems, the
stability of fixed points always switches at saddle-node
bifurcations, and this switching is caused by a linear-stability
eigenvalue of the fixed points crossing zero along the real axis
\cite{Guck_Holm}. For solitary waves in nonlinear partial
differential equations (which can be viewed as fixed points in
infinite-dimensional dynamical systems), it is widely believed that
their stability also always switches at saddle-node bifurcations. We
find that this belief is incorrect. Below, we show that for solitary
waves (\ref{e:Usoliton}) in the generalized NLS equations
(\ref{e:U}), there are no linear-stability eigenvalues crossing zero
at a saddle-node bifurcation point, thus stability-switching does
not occur.

To study the linear stability of solitary waves (\ref{e:Usoliton})
in Eq. (\ref{e:U}), we perturb them as \cite{Yang_SIAM}
\begin{eqnarray}
U(\textbf{x},t)&=&e^{i\mu t} \left\{ u (\textbf{x}
)+[{v}(\textbf{x})+w(\textbf{x})]e^{\lambda t}+ \right. \nonumber \\
&& \hspace{1.5cm} \left. [{v}^*(\textbf{x})-w^*
(\textbf{x})]e^{\lambda^* t} \right\} ,
\end{eqnarray}
where $v, w\ll 1$ are normal-mode perturbations, and $\lambda$ is
the mode's eigenvalue. Inserting this perturbed solution into
(\ref{e:U}) and linearizing, we obtain the following
linear-stability eigenvalue problem
\begin{equation} \label{e:LPhi}
{\cal L}\Phi=-i\lambda \Phi,
\end{equation}
where
\begin{equation}
{\cal L} =  \left[\begin{array}{cc} 0 & L_0 \\ L_1 & 0
\end{array}\right], \quad \Phi = \left[\begin{array}{c} v \\ w
\end{array}\right],
\end{equation}
\begin{eqnarray}
\label{e:L0}
 L_0&=& \nabla^2-\mu+F(u^2, \x),
\end{eqnarray}
and $L_1$ has been given in Eq. (\ref{e:L1}).

At a saddle-node bifurcation point $\mu=\mu_0$, we denote
\begin{equation}
L_{00}=L_0|_{\mu=\mu_0,\ u=u_0}, \quad  {\cal L}_0={\cal
L}_{\mu=\mu_0,\ u=u_0}.
\end{equation}
Then in view of Eq. (\ref{e:u}), we have
\begin{equation}  \label{e:L00u0}
L_{00}u_0=0,
\end{equation}
thus zero is a discrete eigenvalue of $L_{00}$. From this equation
as well as (\ref{e:L10psi}), we have
\begin{equation} \label{e:L0zeromode}
{\cal L}_0 \left[\begin{array}{c} 0 \\ u_0 \end{array}\right]={\cal
L}_0 \left[\begin{array}{c} \psi \\ 0 \end{array}\right]=0,
\end{equation}
thus zero is also a discrete eigenvalue of ${\cal L}_0$.

On the bifurcation of the zero eigenvalue in ${\cal L}_0$ when $\mu$
moves away from $\mu_0$, we have the following main result.

\vspace{0.2cm} \textbf{Theorem 2} \ Assuming that zero is a simple
discrete eigenvalue of $L_{00}$ and $L_{10}$, then at a saddle-node
bifurcation point $\mu_0$, no eigenvalues of the linear-stability
operator ${\cal L}$ cross zero, thus no stability switching occurs.

\vspace{0.2cm} \textbf{Proof.} The idea of the proof is to show
that, when $\mu$ moves away from $\mu_0$, the algebraic multiplicity
of the zero eigenvalue in ${\cal L}$ does not decrease, thus the
zero eigenvalue in ${\cal L}$ cannot bifurcate out to nonzero.

At the saddle-node bifurcation point $\mu=\mu_0$, $(0, u_0)^T$ and
$(\psi, 0)^T$ are two linearly independent eigenfunctions of the
zero eigenvalue in ${\cal L}_0$ in view of Eq. (\ref{e:L0zeromode}).
Here the superscript `T' represents the transpose of a vector. Under
the assumption of Theorem 2, zero is a simple discrete eigenvalue of
$L_{00}$ and $L_{10}$. Thus it is easy to see that ${\cal L}_0$ does
not admit any additional eigenfunctions at the zero eigenvalue,
which means that the geometric multiplicity of the zero eigenvalue
in ${\cal L}_0$ is two. To determine the algebraic multiplicity of
the zero eigenvalue in ${\cal L}_0$, we need to examine the number
of generalized eigenfunctions of this zero eigenvalue. The
lowest-order generalized eigenfunction $(f_1, g_1)^T$ to the
eigenfunction $(0, u_0)^T$ of this zero eigenvalue satisfies the
equation
\begin{equation}
{\cal L}_0 \left[\begin{array}{c} f_1 \\ g_1
\end{array}\right]=\left[\begin{array}{c} 0 \\ u_0
\end{array}\right],
\end{equation}
so the equation for $f_1$ is
\begin{equation} \label{e:L10f1u0}
L_{10}f_1=u_0.
\end{equation}
From Eq. (\ref{e:L10psi}), we see that this inhomogeneous equation
has a homogeneous localized solution $\psi$. In addition, from
conditions of saddle-node bifurcations in Theorem 1, $\langle u_0,
\psi\rangle \ne 0$. Furthermore, $L_{10}$ is a self-adjoint
operator. Thus, from the Fredholm Alternative Theorem, the
inhomogeneous equation (\ref{e:L10f1u0}) does not admit any
localized solution, which means that the eigenfunction $(0, u_0)^T$
of the zero eigenvalue in ${\cal L}_0$ does not have any generalized
eigenfunctions. Similarly, we can show that the eigenfunction
$(\psi, 0)^T$ of the zero eigenvalue in ${\cal L}_0$ does not have
any generalized eigenfunctions either. Hence the algebraic
multiplicity of the zero eigenvalue in ${\cal L}_0$ is equal to its
geometric multiplicity and is two.

Away from the bifurcation point (i.e., $\mu\ne \mu_0$), ${\cal L}$
always has a zero eigenmode
\begin{equation} \label{e:curlL0u0}
{\cal L}\left[\begin{array}{c} 0 \\ u
\end{array}\right]=0
\end{equation}
in view of Eq. (\ref{e:u}). In addition, by differentiating Eq.
(\ref{e:u}) with respect to $\mu$, we also get
\begin{equation}  \label{e:curlL0umu}
{\cal L}\left[\begin{array}{c} u_\mu \\ 0
\end{array}\right]=\left[\begin{array}{c} 0 \\ u
\end{array}\right],
\end{equation}
thus $(u_\mu, 0)^T$ is a generalized eigenfunction of the zero
eigenvalue in ${\cal L}$. This means that the algebraic multiplicity
of the zero eigenvalue in ${\cal L}$ is at least two when $\mu\ne
\mu_0$.

If nonzero eigenvalues bifurcate out from the zero eigenvalue in
${\cal L}$, the algebraic multiplicity of this zero eigenvalue must
decrease. Our results above show that, when $\mu$ moves away from
$\mu_0$, the algebraic multiplicity of the zero eigenvalue in ${\cal
L}$ does not decrease, thus there cannot be nonzero eigenvalues of
${\cal L}$ bifurcating out from zero. Consequently, no eigenvalues
of ${\cal L}$ cross zero at the saddle-node bifurcation point, thus
no stability switching occurs. This completes the proof of Theorem
2.


\vspace{0.1cm} Now we discuss the case when Eq. (\ref{e:U}) is
nonconservative, i.e., the function $F(\cdot, \cdot)$ in (\ref{e:U})
is complex-valued. In this case, if $F$ admits parity-time (PT)
symmetry $F^*(|U|^2, \x)=F(|U|^2, -\x)$, then solitary waves
(\ref{e:Usoliton}) can still exist over a continuous range of real
$\mu$ values \cite{Bender}, and saddle-node bifurcations can also
occur (see later text). By slightly modifying the analysis above, we
can show that there is no stability switching at saddle-node
bifurcations in these nonconservative systems either.

Next we use two examples to confirm the above analytical findings.

\begin{figure}[h!]
\centerline{\includegraphics[width=0.45\textwidth]{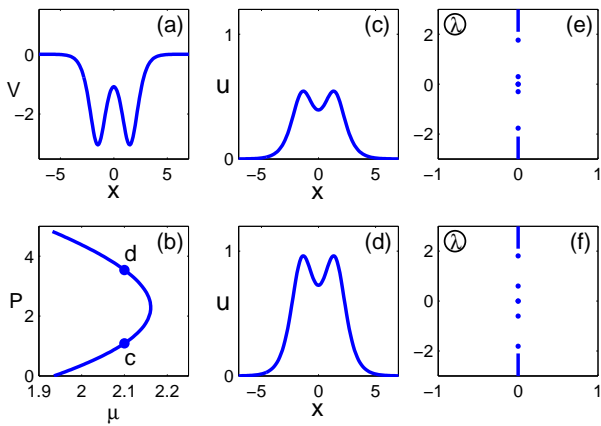}}

\vspace{-0.2cm} \caption{(a) potential (\ref{e:potential1}) in
Example 1; (b) power curve of solitons; (c,d) soliton profiles at
points marked by the same letters in (b); (e,f) stability spectra of
solitons in (c,d). } \label{f:saddle1}
\end{figure}

\textbf{Example 1.} Consider Eq. (\ref{e:U}) with a symmetric
double-well potential and cubic-quintic nonlinearity, i.e.,
\begin{equation}  \label{e:NLScq}
iU_t+U_{xx}-V(x)U+|U|^2U-|U|^4U=0,
\end{equation}
where the double-well potential
\begin{equation}  \label{e:potential1}
V(x)=-3 \left[ \mbox{sech}^2(x+1.5)+ \mbox{sech}^2(x-1.5)\right]
\end{equation}
is shown in Fig. \ref{f:saddle1}(a), and the quintic nonlinearity
has the opposite sign of the cubic nonlinearity. Solitary waves in
this conservative system are of the form (\ref{e:Usoliton}), where
$u(x)$ is real. We have computed these solitary waves by the
Newton-conjugate-gradient method \cite{Yang_SIAM}, and their power
curve is plotted in Fig. \ref{f:saddle1}(b). Here the soliton power
$P$ is defined as $\int_{-\infty}^\infty |u|^2dx$. It is seen that a
saddle-node bifurcation occurs at $\mu_0\approx 2.16$. Two solitary
waves on the lower and upper branches near this bifurcation point
are displayed in Fig. \ref{f:saddle1}(c,d). To determine the linear
stability of these solitary waves, we have computed their whole
linear-stability spectra by the Fourier collocation method
\cite{Yang_SIAM}. These spectra for the two solitary waves in Fig.
\ref{f:saddle1}(c,d) are shown in Fig. \ref{f:saddle1}(e,f)
respectively. It is seen that none of the spectra contains unstable
eigenvalues, indicating that these solitary waves on both lower and
upper branches are linearly stable. We have also performed this
spectrum computation for other solitary waves on the power curve of
Fig. \ref{f:saddle1}(b), and found that they are all linearly
stable. Thus there is no stability switching at the saddle-node
bifurcation point, in agreement with our analytical result.

\begin{figure}[h]
\centerline{\includegraphics[width=0.45\textwidth]{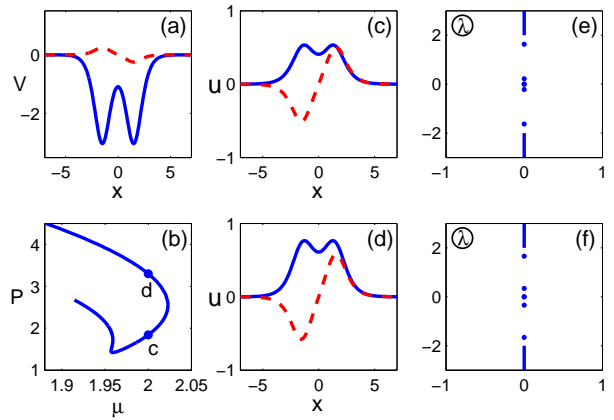}}

\vspace{-0.23cm} \caption{(a) PT potential (\ref{e:PT}) in Example
2; (b) power curve of solitons; (c,d) soliton profiles at points
`c,d' in (b); (e,f) stability spectra of solitons in (c,d). In
(a,c,d), solid line is the real part, and dashed line is the
imaginary part. } \label{f:saddle3}
\end{figure}

\textbf{Example 2.} We still consider Eq. (\ref{e:NLScq}) but now
with a complex PT-symmetric potential
\begin{eqnarray}
V(x) & = & -3\left[ \mbox{sech}^2(x+1.5)+
\mbox{sech}^2(x-1.5)\right] \nonumber
\\ && \hspace{-0.5cm}  +0.25i
\left[\mbox{sech}^2(x+1.5)-\mbox{sech}^2(x-1.5)\right],
\hspace{0.5cm} \label{e:PT}
\end{eqnarray}
see Fig. \ref{f:saddle3}(a). This nonconservative system still
admits solitary waves (\ref{e:Usoliton}) for continuous real ranges
of $\mu$, but $u(x)$ is complex-valued now. We have numerically
obtained a family of these solitons, whose power curve is plotted in
Fig. \ref{f:saddle3}(b). Again a saddle-node bifurcation can be seen
at $\mu_0\approx 2.02$. For solitary waves on the lower and upper
branches near this bifurcation point (see Fig.
\ref{f:saddle3}(c,d)), their stability spectra lie entirely on the
imaginary axis (see Fig. \ref{f:saddle3}(e,f)), indicating that they
are all linearly stable. Hence no stability switching occurs at
saddle-node bifurcations in this nonconservative system either.

\vspace{-0.2cm}

In summary, we have shown that for solitary waves in the generalized
NLS equations (\ref{e:U}) with real or complex potentials, stability
does not switch at saddle-node bifurcations. This disproves a
wide-spread belief that such stability switching should always occur
in nonlinear partial differential equations. Since the generalized
NLS equations (\ref{e:U}) arise frequently in nonlinear optics,
Bose-Einstein condensates and other physical disciplines, our
finding could have broad impact.



\end{document}